\documentclass[12pt]{article}
\usepackage[latin1]{inputenc}
\usepackage{natbib}
\usepackage{amsbsy}
\setcitestyle{comma,aysep={}}           
\newcommand{\dx}[1]{\frac{\partial #1}{\partial \mathbf{x}}}
\newcommand{\lap}[1]{\frac{\partial^2 #1}{\partial \mathbf{x}^2}}
\newcommand{\dxp}[1]{\frac{\partial #1}{\partial \mathbf{x}'}}
\newcommand{\dxpp}[1]{\frac{\partial #1}{\partial \mathbf{x}''}}

\begin{document}

\begin{center}
\Large GENERIC framework for reaction diffusion systems
\end{center}

\begin{center}
\textit{Miguel Hoyuelos} \\ Departamento de F\'{\i}sica, Facultad de Ciencias Exactas y Naturales,\\ Universidad Nacional de Mar del Plata \\ and Instituto de Investigaciones F\'{\i}sicas de Mar del Plata\\ (Consejo Nacional de Investigaciones Cient\'{\i}ficas y T\'{e}cnicas),\\ Funes 3350, 7600 Mar del Plata, Argentina
\end{center}

\begin{abstract}

The GENERIC theory provides a framework for the description of non-equilibrium phenomena in isolated systems beyond local thermal equilibrium and beyond linear non-equilibrium (i.e., linear relations between thermodynamic forces and currents). The framework is applied to some simple and general systems: reactions with detailed balance and Fokker-Planck (FP) equation.  It is shown how to apply the theory to a combination of system and reservoir in the case of the FP equation.
\end{abstract}

\section{Introduction}

A statistical mechanics definition of entropy for systems out of equilibrium remains an open problem despite the efforts devoted to obtain it during many decades, since the days of Boltzmann and Gibbs.
The most common approach, that works in several cases, is a generalization of the Gibbs definition of entropy from equilibrium to non equilibrium systems: $S_G = \int  \rho(t) \ln \rho(t)\, d^n x$, where the time dependent probability density $\rho(t)$ is defined in the $n$-dimensional phase space of micro variables ($\textbf{x}$), but the same form is also used for macroscopic variables.  There are problems with this generalization.  It is well known that $S_G$ remains constant for any Hamiltonian system (see, e.g., \citealt[Sect.\ 5.3]{uffink}), even for time-dependent ensembles describing isolated systems not in equilibrium. \citet{gold-leb} consider that, for non equilibrium systems, the Boltzmann's definition of entropy is more appropriate: $S_B = k_B \ln \Gamma$, where $k_B$ is Boltzmann's constant and $\Gamma$ is the number of micro states consistent with a given macro state.  Regarding the cases where $S_G$ remains constant, \citet{goldstein} says
\begin{quote}
It is frequently asked how this can be compatible with the Second Law. The answer is very simple. The Second Law is concerned with the thermodynamic entropy, and this is given by Boltzmann's entropy, not by the Gibbs entropy.
\end{quote}
But Boltzmann's entropy is not free of problems.  There is some vagueness in its definition. In many cases, it is not clear how to evaluate the number of micro states consistent with a given macro state.

For example, let us consider a linear optical cavity pumped with an external field.  If the evolution is described in terms of the number of photons, it is not difficult to obtain the entropy using Boltzmann's definition.  Furthermore, if $N$ is the number of photons and $M$ is the number of micro states for one photon, it can be shown that, when maximizing $S_B$, one gets Bose-Einstein distribution for $N\sim M$, Boltzmann distribution for $N\ll M$ and the Rayleigh-Jeans limit for $N\gg M$. The situation is more difficult when the description is in terms of the number of photons \textit{and} the phase (including, for example, the effects of detuning or diffraction), because it is not clear how the number of micro states $\Gamma$ depends on the phase.


Because of the difficulties and problems of using the statistical mechanics definitions of entropy in non equilibrium systems, I try to adopt a different approach here. The GENERIC framework, developed in \textit{Beyond Equilibrium Thermodynamics}, \cite{ottinger}, can be used to obtain a form of the entropy function or functional, without considering any a priori definition of entropy from first principles.  The general time-evolution equation for beyond equilibrium systems, or GENERIC, takes a very compact and simple form:
\begin{equation}
\frac{d X}{d t} = L(X) \cdot \frac{\delta E(X)}{\delta X} + M(X) \cdot \frac{\delta S(X)}{\delta X}
\end{equation}
where $X$ is a set of variables that describe a closed system, $E(X)$ and $S(X)$ are the total energy and entropy, and $L(X)$ and $M(X)$ are the Poisson and friction matrices that represent the reversible and irreversible contributions generated by $E(X)$ and $S(X)$ respectively. Operator $\delta/\delta X$ is a partial or functional derivative depending on $E(X)$ and $S(X)$ being functions or functionals of $X$. The Poisson matrix $L(X)$ must be antisymmetric and the friction matrix $M(X)$ must be symmetric and positive semidefinite.  Furthermore, they should satisfy the degeneracy requirements
\begin{eqnarray}
L(X) \cdot \frac{\delta S(X)}{\delta X} &=& 0, \label{degl} \\ 
M(X) \cdot \frac{\delta E(X)}{\delta X} &=& 0. \label{degm}
\end{eqnarray}
These conditions guarantee energy conservation and entropy increase:
\begin{equation}
\frac{dE}{dt} = 0 \ \ \ \ \mbox{and}\ \ \ \ \frac{dS}{dt} \ge 0.
\end{equation}
The conditions on $L$ and $M$ can be transformed into conditions on the derivatives of the entropy, and can be used to determine it  without assuming a specific form from the beginning.  Here, I illustrate this possibility with some examples of systems that, in principle, are not isolated. It is then necessary to include a reservoir in the description, so that the combination of system and reservoir is isolated, and $E(X)$ and $S(X)$ represent the energy and entropy of this combination.  I use some simplifying assumptions in order to keep the description of the combination as simple as possible; they are explained in Sect.\ \ref{assumptions}.

The systems analyzed are:
\begin{itemize}
\item Drift and diffusion system described by a Fokker-Planck equation (Sect.\ \ref{fpeq}).  One particularly interesting case is the Kramers equation, because an always increasing entropy can be obtained for a system that does not necessarily satisfy the local thermal equilibrium assumption.  I show that the conditions on the friction and Poisson matrices can be used to determine that the entropy should belong to a given class of functions.
\item General reaction system with detailed balance (Sect.\ \ref{reactions}). This case can be combined with the previous one to get the GENERIC structure for reaction-diffusion systems.
\item Linear cavity with dispersion and diffraction effects (Sect.\ \ref{cavity}).  Lyapunov functionals for this kind of systems, even with nonlinear terms, are well known (see, e.g., \citealt{mayol} or \citealt{montagne}).  The derivation of the GENERIC structure is presented as an example of the procedure described here for non-isolated systems.
\end{itemize}

The first two examples were already studied by \cite{mielke} using the Wasserstein metric, where a diffusion equation without drift term was used instead of the Fokker-Planck equation. The intention here is to present a different perspective on the same problems.

In Sect.\ \ref{rule} I make some comments on the rule used to determine reversible and irreversible terms.

\section{Simplifying assumptions}
\label{assumptions}

The starting point is a set of equations that determine the evolution of a system in contact with a reservoir.  Variables $\mathbf{n}$ represent the state of the system.  They can be fields depending on the space variable $\mathbf{x}$ of dimension $d$.  The complete description of system and reservoir require additional variables.  Since the initial set $\mathbf{n}$ is restricted to the system, I assume that the total energy density $\epsilon$ is not included in $\mathbf{n}$.  More additional variables may be required to obtain a GENERIC equation; I call them $\mathbf{a}$, and they are still to be specified.  The state variables that describe the combination of system and reservoir are $X = (\mathbf{n},\epsilon, \mathbf{a})$.  The total entropy and energy are $S = \int s\, d^dx$ and $E = \int \epsilon\, d^dx$. The main simplifying assumption is that the entropy density $s$ is a function only of $\mathbf{n}$ and $\mathbf{a}$,
\begin{equation}
s = s(\mathbf{n},\mathbf{a}).
\end{equation}
In other words, the properties of the reservoir are chosen in such a way that the total entropy density is independent of the total energy density.  In general, entropy is a function of a set of extensive variables that include the internal energy $U$, i.e., $S=S(U,\boldsymbol{\mathcal{E}})$, where $\boldsymbol{\mathcal{E}}$ represents the rest of extensive variables; and, if the temperature $T$ is defined, we have that $\frac{\partial S}{\partial U}=1/T$.  I am assuming that there is an equation of state that relates the internal energy with the rest of extensive variables: $U=U(\boldsymbol{\mathcal{E}})$, so that $S$ can be taken as a function  of $\boldsymbol{\mathcal{E}}$ only.  The total energy $E$ is an  independent variable (i.e., knowledge of $\boldsymbol{\mathcal{E}}$ and $U$ is not enough to determine $E$).

Then,
\begin{equation}
\frac{\delta S}{\delta X} = \left( \begin{array}{c} \frac{\partial s}{\partial \mathbf{n}} \\ 0 \\ \frac{\partial s}{\partial \mathbf{a}} \end{array} \right)\ \ \ \mbox{and} \ \ \ 
\frac{\delta E}{\delta X} = \left( \begin{array}{c} \mathbf{0} \\ 1 \\ \mathbf{0} \end{array} \right).
\end{equation}

The available information for the dynamics are the system's evolution equations
\begin{equation}
\frac{\partial \mathbf{n}}{\partial t} = \mathbf{R}(\mathbf{n}),
\end{equation}
where $\mathbf{R}$ is a set of functions that may depend also on space derivatives of $\mathbf{n}$.  From energy conservation, we have that $\int \frac{\partial \epsilon}{\partial t}\, d^dx = 0$.  I also assume that any energy flux in the system is compensated by an opposite energy flux in the reservoir, so that 
\begin{equation}
\frac{\partial \epsilon}{\partial t} = 0.
\end{equation}

What remains to be specified is $\mathbf{a}$ and its evolution.  This is a central point of this paper.  As \citet[p.\ 11]{ottinger} says: ``A crucial part of GENERIC modeling is the choice of the state variables $X$ or, in other words, the definition of a suitable system to describe certain phenomena of interest.'' The criteria are to keep the description as simple as possible and not to add arbitrary information through the evolution of $\mathbf{a}$. In some cases it is not necessary to add any variable $\mathbf{a}$, and it is enough to add only the energy density $\epsilon$.  Some appropriate choices of $\mathbf{a}$, that allow to write GENERIC equations, are proposed in the next sections.

%

\section{The Fokker-Planck equation}
\label{fpeq}

The Fokker-Planck equation is 
\begin{equation}
\frac{\partial n}{\partial t} = - \dx{}\cdot\mathbf{J}\ \ \ \mbox{with}\ \ \  \mathbf{J} = \mathbf{A} n - \frac{1}{2} \dx{}\cdot (B n),
\end{equation}
where $\mathbf{J}$ is the particle current, $\mathbf{A}$ is the drift vector and $B$ is the (positive semidefinite) diffusion matrix.  The space vector $\mathbf{x}$ may include odd variables under time reversal (like momenta).  The system evolves irreversibly to the equilibrium solution $n_e$.  It is convenient to separate the current in their reversible and irreversible parts taking into account that, in equilibrium, the irreversible part should be zero:
\begin{equation}
\mathbf{J} = \underbrace{\mathbf{v} n}_{\mathrm{rev.}} - \underbrace{\frac{n_e}{2} B\cdot \dx{n/n_e}}_{\mathrm{irrev.}},
\end{equation}
with $\mathbf{v}=\mathbf{A} -\frac{1}{2n_e} \dx{}\cdot (Bn_e)$.  According to the rule mentioned in Sect.\ \ref{rule} for determining reversible and irreversible terms, it is necessary to assume that there are physical arguments to state that $\mathbf{v}\rightarrow -\mathbf{v}$ when $t \rightarrow -t$ \footnote{If $\mathbf{x}$ includes odd variables under time reversal, when $t \rightarrow -t$ the space variables change as $\mathbf{x} \rightarrow T\cdot \mathbf{x}$, where $T$ is a diagonal matrix with numbers 1 for even variables and $-1$ for odd variables under time reversal.  In this case, $\mathbf{v}\rightarrow -T\cdot \mathbf{v}$.}.

I propose the following set of variable to describe the combination of system and reservoir: $X = (n,\epsilon, n_e)$.  So, the derivatives of entropy and energy are
\begin{equation}
\frac{\delta S}{\delta X} = \left( \begin{array}{c} \frac{\partial s}{\partial n} \\ 0 \\ \frac{\partial s}{\partial n_e} \end{array} \right)\ \ \ \mbox{and} \ \ \ 
\frac{\delta E}{\delta X} = \left( \begin{array}{c} 0 \\ 1 \\ 0 \end{array} \right).
\end{equation}

The evolution equation is
\begin{equation}
\frac{\partial X}{\partial t} = \left( \begin{array}{c} -\dx{}\cdot (\mathbf{v} n) + \dx{}\cdot \left(\frac{n_e}{2} B\cdot \dx{n/n_e}\right) \\ 0 \\ 0 \end{array} \right).
\end{equation}

\subsection{Friction matrix}

The friction matrix is
\begin{equation}
M(\mathbf{x},\mathbf{x}') = \left(\begin{array}{ccc} M_{11}(\mathbf{x},\mathbf{x}') & 0 & 0 \\
0 & 0 & 0 \\
0 & 0 & 0 \end{array} \right),
\end{equation}
it satisfies the degeneracy condition (\ref{degm}) and it should also satisfy
\begin{equation}
\left.\frac{\partial X}{\partial t}\right|_\mathrm{irrev} = M(\mathbf{x},\mathbf{x}')\bullet \frac{\delta S}{\delta X} = \left( \begin{array}{c} \dx{}\cdot \left(\frac{n_e}{2} B\cdot \dx{n/n_e}\right) \\ 0 \\ 0 \end{array} \right),
\end{equation} 
The notation of \citet{ottinger} was used, where the big dot $\bullet$ means matrix multiplication and integration in the continuous index $\mathbf{x}'$, and $\frac{\delta S}{\delta X}$ is evaluated in $\mathbf{x}'$. Then
\begin{equation}
\int M_{11}(\mathbf{x},\mathbf{x}') \left.\frac{\partial s}{\partial n}\right|_{\mathbf{x}'}\, d^d x' = \dx{}\cdot \left(\frac{n_e}{2} B\cdot \dx{n/n_e}\right),
\end{equation}
or, more generally,
\begin{equation}
\int M_{11}(\mathbf{x},\mathbf{x}') \left.\frac{\partial s}{\partial n}\right|_{\mathbf{x}'}\, d^d x' = \dx{}\cdot \left(\frac{n_e}{2f'} B\cdot \dx{f}\right),
\label{m110}
\end{equation}
where $f$ is a function of $n/n_e$.  The solution for $M_{11}(\mathbf{x},\mathbf{x}')$ is
\begin{equation}
M_{11}(\mathbf{x},\mathbf{x}') = \frac{f(\mathbf{x}')}{ \left.\frac{\partial s}{\partial n}\right|_{\mathbf{x}'}}  \dxp{}\cdot \left[ \frac{n_e(\mathbf{x}')}{2f'(\mathbf{x}')} B(\mathbf{x}') \cdot \dxp{\delta(\mathbf{x}-\mathbf{x}')} \right],
\label{m11}
\end{equation}
where $f(\mathbf{x}')=f[n(\mathbf{x}')/n_e(\mathbf{x}')]$ and $\left.\frac{\partial s}{\partial n}\right|_{\mathbf{x}'}$ means that this derivative is evaluated in $(n(\mathbf{x}'),n_e(\mathbf{x}'))$. It can be shown that (\ref{m11}) is solution by replacing it in (\ref{m110}), integrating twice by parts and neglecting boundary terms.

$M_{11}(\mathbf{x},\mathbf{x}')$ must be positive semidefinite, i.e., it should be possible to write it as\footnote{The condition for a matrix $M$ to be positive semidefinite is that it can be written as $M= C\cdot C^T$.  The generalization of this condition to continuous indexes is $M(\mathbf{x},\mathbf{x}') = \int d^d x'' V(\mathbf{x},\mathbf{x}'') V^T(\mathbf{x}'',\mathbf{x}') = \int d^d x'' V(\mathbf{x},\mathbf{x}'') V(\mathbf{x}',\mathbf{x}'')$. In (\ref{positive}), $V$ was replaced by vector $\mathbf{V}$ using the fact that the sum of positive semidefinite matrices is also positive semidefinite.}
\begin{equation}
M_{11}(\mathbf{x},\mathbf{x}') = \int d^d x''\, \mathbf{V}(\mathbf{x},\mathbf{x}'')\cdot \mathbf{V}(\mathbf{x}',\mathbf{x}''), \label{positive}
\end{equation}
where $\mathbf{V}(\mathbf{x},\mathbf{x}'')$ is a vector of, in principle, arbitrary dimension.  To get this form, we rewrite (\ref{m11}) as
\begin{eqnarray}
M_{11}(\mathbf{x},\mathbf{x}') &=& \int d^d x''\, \delta(\mathbf{x}'-\mathbf{x}'') M_{11}(\mathbf{x},\mathbf{x}'') \nonumber \\
&=& - \int d^d x''\, \dxpp{} \left[\delta(\mathbf{x}'-\mathbf{x}'') \frac{f(\mathbf{x}')}{ \left.\frac{\partial s}{\partial n}\right|_{\mathbf{x}'}} \right]\cdot \left[\frac{n_e(\mathbf{x}'')B(\mathbf{x}'')}{2f'(\mathbf{x}'')} \right] \cdot \dxpp{\delta(\mathbf{x}-\mathbf{x}'')}.
\end{eqnarray}
The only factor that contains $\mathbf{x}$ is $\dxpp{\delta(\mathbf{x}-\mathbf{x}'')}$, and it should be included in $\mathbf{V}(\mathbf{x},\mathbf{x}'')$.  This means that there should be also a factor $\dxpp{\delta(\mathbf{x'}-\mathbf{x}'')}$ for $\mathbf{V}(\mathbf{x'},\mathbf{x}'')$.  Taking 
\begin{equation}
\frac{\partial s}{\partial n} = - f
\end{equation}
we obtain the desired form with
\begin{equation}
\mathbf{V}(\mathbf{x},\mathbf{x}'') = \sqrt{\frac{n_e(\mathbf{x}'')}{2f'(\mathbf{x}'')}}\, C \cdot \dxpp{\delta(\mathbf{x}-\mathbf{x}'')},
\label{v}
\end{equation}
where I have used the fact that $B=C^T\cdot C$ is positive semidefinite. From the square root in (\ref{v}) we have that $f$ should satisfy the condition $f'>0$.

In summary, from the conditions for the friction matrix we obtained that $\frac{\partial s}{\partial n}$ is equal to minus any function of $n/n_e$ with positive derivative.  So, the entropy density is 
\begin{equation}
s(n,n_e) = -n_e F(n/n_e) + F_e(n_e),
\label{entr1}
\end{equation}
where $F'=f$, $F''>0$, and $F_e$ is a function of $n_e$.

Using the maximum condition we have that $\left.\frac{\partial s}{\partial n} \right|_{n=n_e} = -F'(1) = 0$.  Let us note that, if the constraint $\int n(\mathbf{x}) \,d^d x = \mathrm{const.}$ is considered, an arbitrary constant may be added to $F'$ (see \citealt[p.\ 454]{ottinger}).

\subsection{Poisson matrix}

The Poisson matrix is 
\begin{equation}
L(\mathbf{x},\mathbf{x}') = \left( \begin{array}{ccc} 0 & n(\mathbf{x}') \mathbf{v}(\mathbf{x}')\cdot \dxp{\delta} & 0 \\
-n(\mathbf{x}) \mathbf{v}(\mathbf{x})\cdot \dx{\delta} & 0 & -n_e(\mathbf{x}) \mathbf{v}(\mathbf{x})\cdot \dx{\delta} \\
0 & n_e(\mathbf{x}') \mathbf{v}(\mathbf{x}')\cdot \dxp{\delta} & 0 
\end{array} \right)
\end{equation}
where $\delta = \delta(\mathbf{x}-\mathbf{x}')$.  It is easy to see that the reversible part of the dynamics is given by
\begin{equation}
\left. \frac{\partial X}{\partial t} \right|_\mathrm{rev.} = L(\mathbf{x},\mathbf{x}') \bullet \frac{\delta E}{\delta X} = \left( \begin{array}{c} -\dx{}\cdot (\mathbf{v} n) \\ 0 \\ 0 \end{array} \right).
\label{poissonfp}
\end{equation}

From the degeneracy condition (\ref{degl}) we obtain
\begin{equation}
n\mathbf{v}\cdot \dx{} \frac{\partial s}{\partial n} + n_e\mathbf{v}\cdot \dx{} \frac{\partial s}{\partial n_e} = 0.
\label{degcond}
\end{equation}
From (\ref{entr1}) we have that $\frac{\partial s}{\partial n} = -F'$ and $\frac{\partial s}{\partial n_e} = -F + \frac{n}{n_e} F' + F_e'$, and, replacing in (\ref{degcond}), we get $F_e'' = 0$.  Then, $F_e(n_e) = c_1 n_e + c_2$; constant $c_2$ can be ignored (it represents a shift in the origin of the entropy) and the term $c_1 n_e$ can be absorbed into $F$, so that
\begin{equation}
s = -n_e F(n/n_e)
\label{entr}
\end{equation}
with $F'(1)=0$ and $F''(n/n_e) >0$. Eq.\ (\ref{entr}) defines a class of functions for the entropy.  The function $s = -n \ln(n/n_e) + n$ is one member of the class.

All members of the class have a similar behavior close to equilibrium
\begin{equation}
s \simeq - \frac{(n-n_e)^2}{n_e}
\end{equation}
where it was assumed that $F(1)=0$, and $F''(1)$ was absorbed in the units of $s$.

We got a GENERIC structure for the Fokker-Planck equation, with matrices $M$ and $L$ that satisfy all required conditions.  The conditions were used to show that the entropy should belong to a determined class of functions.

\section{Reactions}
\label{reactions}

Let us consider a homogeneous system of $m$ reacting species, with densities $\mathbf{n} = (n_1,...,n_m)$.  
There are $r$ possible reactions, with stoichiometric coefficients $\boldsymbol{\alpha}_j$ ($\boldsymbol{\beta}_j$) for the forward (backward) reaction number $j$.  The reaction rate is
\begin{equation}
J_j = k_{+j} \mathbf{n}^{\boldsymbol{\alpha}_j} - k_{-j} \mathbf{n}^{\boldsymbol{\beta}_j} \quad \mbox{for}\ j=1,...,r
\end{equation}
where $\mathbf{n}^{\boldsymbol{\alpha}_j} = n_1^{\alpha_{1j}}...n_m^{\alpha_{mj}}$, and $k_{+j}$ and $k_{-j}$ are the forward and backward reaction constants.

The evolution equation for the densities is
\begin{equation}
\frac{\partial \mathbf{n}}{\partial t} = \mathbf{R}(\mathbf{n}).
\label{evolreac}
\end{equation}
The reaction terms are given by 
\begin{equation}
\mathbf{R} = \nu \cdot \mathbf{J}
\end{equation}
where $\nu$ is a $m\times r$ matrix of stoichiometric coefficients, with components $\nu_{ij} = \beta_{ij} - \alpha_{ij}$.  I assume that detailed balance is satisfied, i.e., there is a stationary solution $\mathbf{n}_e = (n_{e1},...,n_{em})$ for which $\mathbf{J}(\mathbf{n}_e) = \mathbf{0}$.  Using this condition, we can rewrite the reaction rates as
\begin{equation}
J_j = c_j \left( \frac{\mathbf{n}^{\boldsymbol{\alpha}_j}} {\mathbf{n}_e^{\boldsymbol{\alpha}_j}} -  \frac{\mathbf{n}^{\boldsymbol{\beta}_j}} {\mathbf{n}_e^{\boldsymbol{\beta}_j}} \right),
\end{equation}
with $c_j = k_{+j} \mathbf{n}_e^{\boldsymbol{\alpha}_j} = k_{-j} \mathbf{n}_e^{\boldsymbol{\beta}_j} \ge 0$.

For a non-homogeneous system with diffusion processes, we have to use variables $X=(\mathbf{n},\epsilon ,\mathbf{n}_e)$ and generalize the result of the previous section to a multicomponent system.  For a homogeneous system, it is possible to derive the GENERIC structure without additional variables, i.e., we can use $X=(\mathbf{n},\epsilon)$ for the description of system and reservoir.

The evolution is irreversible and there are not reversible terms in (\ref{evolreac}), since if time is inverted ($t\rightarrow -t$) none of the terms in the right hand side changes sign.
This means that the Poisson matrix $L$ is equal to 0.  The derivatives of entropy and energy are
\begin{equation}
\frac{\partial S}{\partial X} = \left( \begin{array}{c} \frac{\partial S}{\partial \mathbf{n}} \\ 0 \end{array} \right)\ \ \ \mbox{and} \ \ \ 
\frac{\partial E}{\partial X} = \left( \begin{array}{c} \mathbf{0} \\ 1 \end{array} \right).
\end{equation}

The friction matrix is 
\begin{equation}
M = \left( \begin{array}{cc} \frac{\mathbf{R}\mathbf{R}}{\sigma} & \mathbf{0} \\ \mathbf{0} & 0 \end{array} \right),
\end{equation}
where $\sigma = \mathbf{R}\cdot \frac{\partial S}{\partial \mathbf{n}}$ and $\mathbf{R}\mathbf{R}$ is a dyadic tensor with components $[\mathbf{R}\mathbf{R}]_{ij}=R_iR_j$.  It is easy to see that 
\begin{equation}
\frac{\partial X}{\partial t} = M\cdot \frac{\partial S}{\partial X} = \left( \begin{array}{c} \mathbf{R} \\ 0 \end{array} \right),
\end{equation}
and that $\sigma$ is equal to the entropy production rate $\frac{\partial S}{\partial t}$.  It is also easy to see that $M$ satisfies the degeneracy condition (\ref{degm}). Matrix $\mathbf{R}\mathbf{R}$ is positive semidefinite; it can be written as $\mathbf{R}\mathbf{R}=C \cdot C^T$, where $C=\mathbf{R}\mathbf{1}$, with $\mathbf{1}$ being a $m$-dimensional unit vector.  Therefore, the only condition to have $M$ positive semidefinite is $\sigma \ge 0$.

It is known that 
\begin{equation}
S = \sum_i [n_i - n_i \ln(n_i/n_{ei})]
\end{equation}
satisfies $\sigma \ge 0$ for a reaction system with detailed balance \citep{mielke}.  This can be proved as follows:
\begin{eqnarray}
\sigma & = & \mathbf{J}\cdot \nu^T \cdot \frac{\partial S}{\partial \mathbf{n}} \nonumber \\
&=& -\sum_{i,j} c_j \left( \frac{\mathbf{n}^{\boldsymbol{\alpha}_j}} {\mathbf{n}_e^{\boldsymbol{\alpha}_j}} -  \frac{\mathbf{n}^{\boldsymbol{\beta}_j}} {\mathbf{n}_e^{\boldsymbol{\beta}_j}} \right) \,
(\beta_{ji} - \alpha_{ji}) \, \ln\frac{n_i}{n_{ei}} \nonumber \\
&=& \sum_j c_j (a_j - b_j) \ln \frac{a_j}{b_j} \ge 0
\end{eqnarray}
where $i=1...m$, $j=1...r$, $a_j = \mathbf{n}^{\boldsymbol{\alpha}_j}/\mathbf{n}_e^{\boldsymbol{\alpha}_j}$ and $b_j = \mathbf{n}^{\boldsymbol{\beta}_j}/ \mathbf{n}_e^{\boldsymbol{\beta}_j}$.

\section{Linear cavity with dispersion and diffraction}
\label{cavity}

In this section I analyze a single-mode cavity with a linearly polarized 
electromagnetic field.  Non linear effects are neglected.  The equation for the complex amplitude $A$ is
\begin{equation}
\frac{\partial A}{\partial t} = A_\mathrm{in} - (\gamma + i\theta) A + (\alpha + i\beta) \lap{A},
\end{equation}
where $A_\mathrm{in}$ is the pump field, $\gamma\ge 0$ is the cavity decay rate, $\theta$ is the cavity detuning, $\alpha\ge 0$ represents dispersion, $\beta$ is diffraction, and $\lap{}$ is the transverse Laplacian (in the plane perpendicular to light propagation).  

The change of variables $a = A-A_s$ is proposed, where $A_s$ is a stationary solution ($A_s$ and $A_\mathrm{in}$ may depend on space).  The equation for $a$ is
\begin{equation}
\frac{\partial a}{\partial t} =  - (\gamma + i\theta) a + (\alpha + i\beta) \lap{a}.
\end{equation}
The complex variable $a$ in terms of its real an imaginary parts is $a=a_1 + i a_2$.  The real part $a_1$ is even under time inversion, while the imaginary part $a_2$ is odd.  Under a time inversion, the terms with $\theta$ and $\beta$ remain unchanged, they are reversible.  The terms with $\gamma$ and $\alpha$ change sign when time is inverted, they are irreversible.

The variables proposed to describe the problem are $X=(a_1, a_2, \epsilon)$. The derivatives of entropy and energy are
\begin{equation}
\frac{\delta S}{\delta X} = \left( \begin{array}{c} \frac{\partial s}{\partial a_1}\\ \frac{\partial s}{\partial a_2} \\ 0  \end{array} \right)\ \ \ \mbox{and} \ \ \ 
\frac{\delta E}{\delta X} = \left( \begin{array}{c} 0 \\ 0 \\ 1  \end{array} \right).
\end{equation}
The evolution equations are
\begin{equation}
\frac{\partial X}{\partial t} = \left( \begin{array}{c} 
-\gamma a_1 + \theta a_2 + \alpha \lap{a_1} - \beta \lap{a_2} \\
-\gamma a_2 - \theta a_1 + \alpha \lap{a_2} + \beta \lap{a_1} \\
0 
\end{array} \right).
\end{equation}
The reversible part of the dynamics is given by 
\begin{equation}
\left.\frac{d X}{d t}\right|_\mathrm{rev.} = L(\mathbf{x},\mathbf{x}') \bullet \frac{\delta E}{\delta X}
\end{equation}
with
\begin{equation}
L(\mathbf{x},\mathbf{x}') = \left( \begin{array}{ccc}
0 & 0 & \theta a_2 \delta + \beta \dxp{a_2} \cdot \dxp{\delta}  \\
0 & 0 & -\theta a_1 \delta - \beta \dxp{a_1} \cdot \dxp{\delta}  \\
-\theta a_2 \delta \atop - \beta \dx{a_2}\cdot \dx{\delta} & \theta a_1 \delta \atop + \beta \dx{a_1}\cdot \dx{\delta} & 0 
\end{array} \right)
\end{equation}
where, as before, $\delta = \delta(\mathbf{x}-\mathbf{x}')$.

From the degeneracy condition (\ref{degl}) we obtain
\begin{equation}
-\theta a_2 \frac{\partial s}{\partial a_1} + \beta \dx{a_2}\cdot \dx{} \frac{\partial s}{\partial a_1} + \theta a_1 \frac{\partial s}{\partial a_2} - \beta \dx{a_1}\cdot \dx{} \frac{\partial s}{\partial a_2} = 0.
\end{equation}
A solution for this equation is $s=c_1(a_1^2 + a_2^2) + c_2$.  Constant $c_2$ can be ignored, and, as we will see in the next paragraphs, $c_1$ must be negative in order to have a positive semidefinite friction matrix.  I use following expression for the entropy density:
\begin{equation}
s = -\frac{1}{2}(a_1^2 + a_2^2).
\end{equation}

The irreversible part of the dynamics is given by 
\begin{equation}
\left.\frac{d X}{d t}\right|_\mathrm{irrev.} = M(\mathbf{x},\mathbf{x}') \bullet \frac{\delta S}{\delta X},
\end{equation}
with
\begin{equation}
\frac{\delta S}{\delta X} = \left( \begin{array}{c} -a_1 \\ -a_2 \\ 0  \end{array} \right)
\end{equation}
and the friction matrix $M$ is given by
\begin{equation}
M(\mathbf{x},\mathbf{x}') = \left( \begin{array}{ccc}
\gamma - \alpha \lap{\delta} & 0 & 0  \\
0 & \gamma - \alpha \lap{\delta} & 0  \\
0 & 0 & 0 \end{array} \right).
\end{equation}
The matrix $M(\mathbf{x},\mathbf{x}')$ is symmetric, positive semidefinite, and satisfies the degeneracy condition (\ref{degm}).

\section{Comments on the rule to determine reversible and irreversible terms}
\label{rule}

The rule generally used to determine if an equation is reversible or irreversible is based on the time reversed ($t\rightarrow -t$) equation.  If the same equation is obtained, then it is reversible, if a minus sign appears, it is irreversible.  Hamilton equations of classical dynamics are reversible; the Schrödinger equation is also reversible (for complex quantities, time inversion also implies complex conjugation).  The diffusion equation is irreversible.  An equation can have a mixture of reversible and irreversible terms, those that preserve sign under time inversion are reversible, and those that change sign are irreversible.

This is the rule used in the previous examples, and it generally works, but not always.  There are cases where the irreversible behavior is manifested not in the differential equations but in the initial conditions, and the rule refers only to differential equations.
The following simple example is useful to illustrate this problem.  Let us consider the Newton's cooling equation
\begin{equation}
\frac{dx}{dt} = - a x,
\end{equation}
where $x$ is the temperature difference between two bodies and $a>0$ is related to heat conductivity.  The equation is obviously irreversible, and entropy is produced due to the heat flux from one body to the other.  Now let us consider the equivalent system
\begin{eqnarray}
\frac{dx}{dt} &=& - a x y \nonumber \\
\frac{dy}{dt} &=& 0
\end{eqnarray}
where $y$ is defined as odd under time reversal, and it has initial condition $y_0=1$. According to the rule, the system is now reversible, and it has zero entropy production.  This is wrong; the rule does not work in this case.  The irreversibility is still there, in the allowed values for the initial condition of $y$.  When we invert time, we get exactly the same differential equations, but, since $y$ is odd under time reversal, the initial condition is now $y_0=-1$, and this value violates the second law of thermodynamics.

The argument is very similar to the one used by Boltzmann in a discussion with Ostwald regarding the irreversibility paradox (cited in \citealt[replies p.\ 115]{lebowitz}): 
\begin{quote}
From the fact that the differential equations of mechanics are left unchanged by reversing the sign of time without anything else, Herr Ostwald concludes that the mechanical view of the world cannot explain why natural processes run preferentially in a definite direction.  But such a view appears to me to overlook that mechanical events are determined not only by differential equations, but also by initial conditions.
\end{quote}

In summary, I think that the rule to distinguish reversible and irreversible terms should be revised and take into account, in some way, the initial conditions.

\section{Final remarks}

I presented the GENERIC equation for three examples (Fokker-Planck equation, reactions and linear cavity) and, using the conditions on the Poisson and friction matrices, obtained an expression for the entropy in each case.

The first difficulty present in the examples is that they consist in non-isolated systems, and the GENERIC equation (in its basic formulation) applies to isolated systems. The approach proposed here is to extend the description to system and reservoir including additional variables, so that the combination is isolated. It is explained, in Sect.\ \ref{assumptions}, how to do this extension using some simplifying assumptions. The main assumptions are that the entropy density is independent of the energy density $\epsilon$, and that the energy density does not change with time (the energy density of the system may change, but not the total one).  Besides $\epsilon$, there are the additional variables $\mathbf{a}$ to be specified for each system.  For the Fokker-Planck equation, the equilibrium density $n_e$ was used as the additional variable, with  $\frac{\partial n_e}{\partial t} = -\dx{}\cdot (\mathbf{v} n_e) = 0$. The inclusion of $n_e$ in the description makes it possible to satisfy the degeneracy condition (\ref{degcond}) of the Poisson matrix (\ref{poissonfp}).  For the homogeneous reaction system and the linear cavity it was enough to add only the energy density. 

The method proposed is, of course, not systematic, since the election of $\mathbf{a}$ depends on the specific system.  But the guidelines illustrated with these examples may be relevant to a wide class of systems.

\end{document}